# Status of Superconducting RF Linac Development for APT

K. C. D. Chan, B. M. Campbell, D. C. Gautier, R. C. Gentzlinger, J. G. Gioia, W. B. Haynes, D. J. Katonak, J. P. Kelley, F. L. Krawczyk, M. A. Madrid, R. R. Mitchell, D. I. Montoya, E. N. Schmierer, D. L. Schrage, A. H. Shapiro, T. Tajima, J. A. Waynert, LANL, Los Alamos, NM 87544, USA

J. Mammosser, TJNAF, Newport News, VA 23606, USA

J. Kuzminski, General Atomics, San Diego, CA 92186, USA

*Abstract*

This paper describes the development progress of high-current superconducting RF linacs in Los Alamos, performed to support a design of the linac for the APT (Accelerator Production of Tritium) Project. The APT linac design includes a CW superconducting RF high-energy section, spanning an energy range of 211–1030 MeV, and operating at a frequency of 700 MHz with two constant- sections ( = 0.64 and =0.82). In the last two years, we have progressed towards building a cryomodule with =0.64. We completed the designs of the 5-cell superconducting cavities and the 210-kW power couplers, and are currently testing the cavities and the couplers. We are scheduled to begin assembly of the cryomodule in September 2000. In this paper, we present an overview of the status of our development efforts and a report on the results of the cavity and coupler testing program.

## 1 INTRODUCTION

Development of Superconducting (SC) RF Technology for high-current CW proton linacs, performed in support of the linac design for the APT Project [1], has been underway at Los Alamos since 1997 [2]. The goal is to design, build and test 5-cell cavities and power couplers to their specifications and integrate them into a prototype cryomodule. Since our report given at the 19th International Linac Conference [3] in August 1998, we finished the design of the cavities and couplers, which were built and tested. We also completed the design of the prototype cryomodule. The major components of the cryomodule are now being fabricated.

In this paper, we will report the test results of the 5-cell cavities, the test results of the power couplers, and the progress of cryomodule fabrication. We will also report the results of a series of tests at cryogenic temperature that determined experimentally the heat leaks between the 2-K operating temperature of the cavities and room temperature. Because of space limitation, we will not describe the designs of the cavities, the couplers, or the cryomodule, as they can be found in Ref. 3.

## 2 STATUS OF CAVITY DEVELOPMENT

The APT 5-cell cavities consist of the bare cavities made of niobium (RRR=250), and the inner and outer helium vessels made of unalloyed titanium [4]. A completed bare cavity is shown in Fig. 1. These bare cavities were fabricated by CERCA in France. The half-cells were formed by spinning, and they were electron-beam welded to form bare cavities. The fabricated bare cavities are RF-tested before they are sent to Titanium Fabrication Corporation for installation of an inner helium vessel by an electron-beam-welding process. To date, inner helium vessels have been installed on two of the four cavities (Fig. 2). Following inner-helium-vessel installation, the cavities will be RF tested again to insure that cavity performance has not been degraded during the inner-helium-vessel installation. After the required performance is confirmed, outer helium vessels, instrumentation, and tuners will be installed to complete the cavities for integration into the cryomodule.

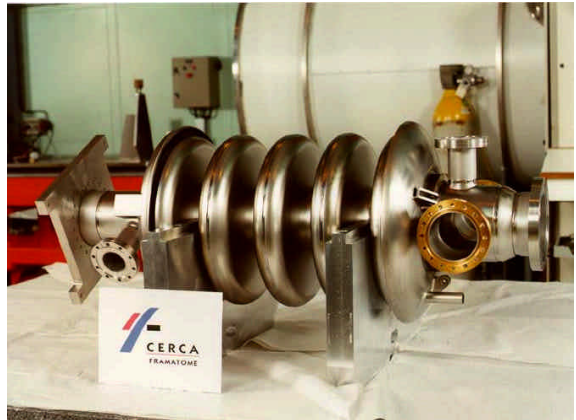

Figure 1  Bare niobium 5-cell cavity

Cavity testing is being shared between Los Alamos National Laboratory (LANL) and Thomas Jefferson National Accelerator Facility (TJNAF). To date, a total of three cavities have been tested successfully.

Before testing, cavity processing includes the following: 1) removal of 150 μm with (1,1,2) buffered-chemistry polishing at acid temperatures below 15˚C; 2)

---

Work supported by US Department of Energy

high-pressure water rinsing; and 3) cavity baking at 150°C.

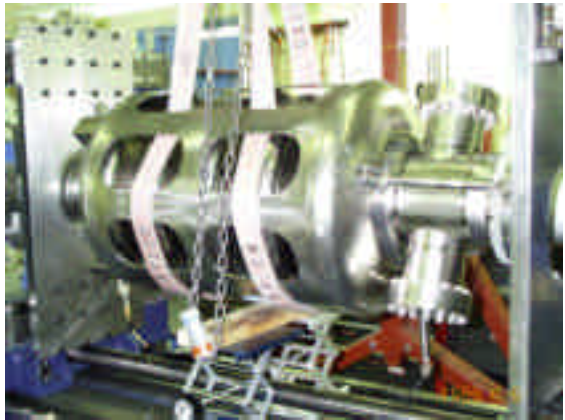

Figure 2  Cavity with inner helium vessel installed

Caviy test results so far are summarized in Fig. 3. The Q-values achieved have been more than a factor of two higher than the required $5\times10^9$ at the operating gradient of 5 MV/m. The highest gradient reached of 10 MV/m is a factor of two higher than the operating gradient. For these cavities, we observed electron activities starting at 3 MV/m. The highest gradients were achieved with helium processing. After helium processing, cavity performance shown remained if the cavities were kept at LHe temperature. The highest gradient reached was limited by excess radiation and/or not enough RF power, and not by quenches.

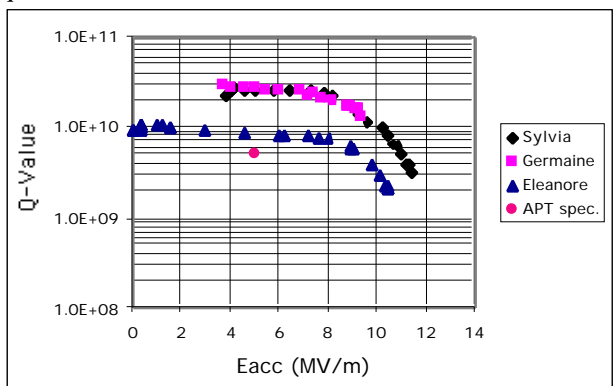

Figure 3  Performance of APT 5-cell cavities

## 3  STATUS OF COUPLER DEVELOPMENT

The APT power couplers are required to transmit 210 kW of RF power [5]. They were tested on a test stand at room temperature (Fig. 4) [6]. Three types of tests were performed: transmitted power, totally reflected power, and condensed gas. The first test, transmitted-power capability, was performed simultaneously with two power couplers up to 1-MW. Power was transmitted from one coupler, through a copper pillbox cavity, to a second coupler, and finally to a water-cooled RF load. The couplers and copper cavity were matched for 100% transmission. The second test, totally-reflected power was performed by replacing the RF load with a "sliding short" and completely reflecting forward power, thus setting up a standing-wave pattern in the coupler-cavity system. By varying the location of the sliding short, we were able to locate the maximum of the standing-wave pattern at sixteen locations over one wavelength along the system. This test allowed us to simulate the situation when the CW beam is interrupted. The third test, the condensed-gas test, was performed by cooling the tapered end of the outer conductors to LN temperature, allowing observation of effects of residual-gas condensation. Condensed-gas effects were reported to be important in enhancing multipacting.

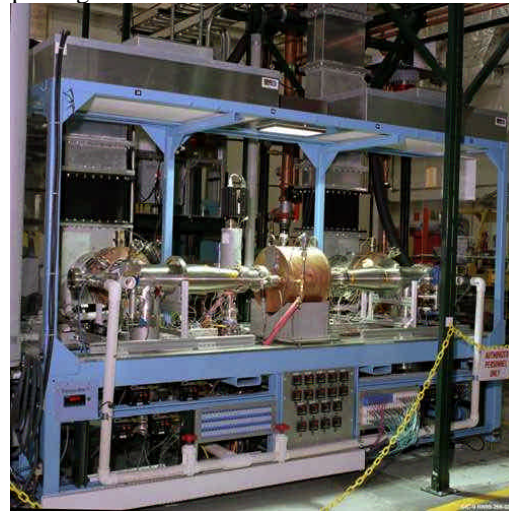

Figure 4  Test stand for coupler at room temperature

Couplers with fixed and adjustable coupling were tested. The adjustability of the coupler was accomplished with BeCu bellows (tip bellows) located close to the tip of the inner conductor (Fig. 5). RF power was fed through RF windows with two ceramic disks. These windows were fabricated by EEV, England, and were tested to 1 MW.

Results of high-power testing are summarized here:
1. Both fixed and adjustable couplers achieved a power level of 1 MW CW in the transmitted-power tests;
2. In the reflected-power tests, the fixed couplers achieved 850 kW CW at the APT operating condition; achieved 550 kW CW and 850 kW at 50% duty cycle at all sliding short positions;
3. We did not observe any significant multipacting in any of the tests. With the couplers operating at $10^{-7}$ Torr, there was some vacuum activity that changed the residual gas pressures at levels of $10^{-9}$ Torr around 250 kW;
4. Data from the condensed-gas tests did not show any enhancement of multipacting

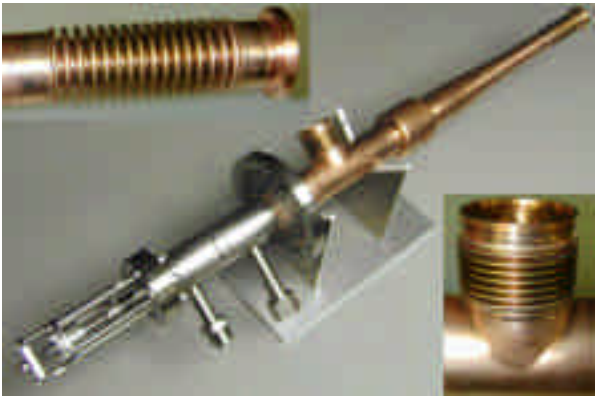

Figure 5  Adjustable inner conductor

After high-power operations of the adjustable couplers, we observed failures of the tip bellows and electron etching. Two tip bellows failed after reaching 750 kW. Preliminary inspection indicated that the failures were caused by excess temperature at the tip bellows, which agreed with temperatures as calculated in thermal simulations. Thermal simulations also showed that the temperature of the tip-bellows could be effectively reduced by 400°C by copper plating the BeCu bellows and cooling using gaseous helium (instead of air) as in the cryomodule design. The electron etching is marks on the inner conductors that look like electron tracks. The density of the etching and the depths of the marks increase with RF power levels.

The coupler test results have been encouraging. The APT Project Office has initiated a change in the coupler specification from 210 kW to 420 kW. This change decreases the number of couplers by one-half, leading to a cost saving of more than $60M in the APT plant design.

## 4  STATUS OF CRYOMODULE DEVELOPMENT

We have completed the final design of the cryomodule [7]. Fabrication of major components is in progress at Ability Engineering Technology in Chicago (Fig.6). All the components will be delivered to Los Alamos by the middle of September.

## 5  CRYOGENIC TEST RIG RESULTS

We completed the experiments that measured the heat leaks from room temperature to 2 K via a power coupler [8]. In our design, this heat leak is minimized by a double-point heat-intercept approach. Without simulated RF heating, the heat leak to 2 K by one coupler is measured as one watt. The low-temperature and high-temperature heat intercept, respectively, removed 2.5 watts and 12 watts of heat. With RF heating simulated for 210 kW of coupler transmitted power, the heat leak to 2 K was 1.4 watt. The heat removed by the low-temperature and high-temperature heat exchangers was, respectively, 10 and 30 watts. This performance is as predicted by our thermal model.

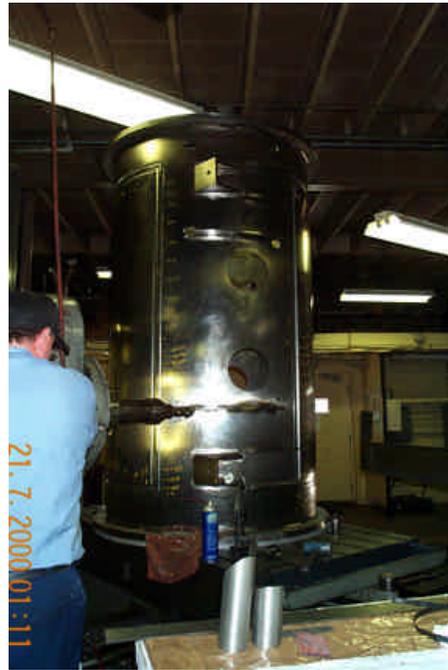

Figure 6  Milling the o-ring grove on the cryostat

## 6  SUMMARY

The SCRF development has tested the 5-cell cavities and couplers experimentally. The measured RF and thermal performance greatly exceeds the APT required performance. We will be ready to assemble the cryomodule in September 2000.